\documentclass[fleqn,10pt]{wlscirep}
\usepackage[utf8]{inputenc}
\usepackage[T1]{fontenc}
\usepackage{subfigure}
\usepackage{lineno}

\title{Evidence for Three-$\alpha$ Breathing Modes Uncovered by Control Neural Network}

\author[1,2]{Zheng Cheng}
\author[1,2,*]{Mengjiao Lyu}
\author[3,4]{Takayuki Myo}
\author[4]{Hisashi Horiuchi}
\author[4]{Hiroshi Toki}
\author[5,6]{Zhongzhou Ren}
\author[7]{Masahiro Isaka}
\author[1,2]{Mengyun Mao}
\author[8]{Hiroki Takemoto}
\author[9]{Niu Wan}
\author[1,2]{Wenlong You}
\author[10]{Qing Zhao}
\affil[1]{College of Physics, Nanjing University of Aeronautics and Astronautics, Nanjing 210016, China}
\affil[2]{Key Laboratory of Aerospace Information Materials and Physics (NUAA), MIIT, Nanjing 211106, China}
\affil[3]{General Education, Faculty of Engineering, Osaka Institute of Technology, Osaka, Osaka 535-8585, Japan}
\affil[4]{Research Center for Nuclear Physics (RCNP), Osaka University, Osaka 567-0047, Japan}
\affil[5]{School of Physics Science and Engineering, Tongji University, Shanghai 200092, China}
\affil[6]{Key Laboratory of Advanced Micro-Structure Materials, Ministry of Education, Shanghai 200092, China}
\affil[7]{Hosei University, 2-17-1 Fujimi, Chiyoda-ku, Tokyo 102-8160, Japan}
\affil[8]{Faculty of Pharmacy, Osaka Medical and Pharmaceutical University, Takatsuki, Osaka 569-1094, Japan}
\affil[9]{School of Physics and Optoelectronics, South China University of Technology, Guangzhou 510641, China}
\affil[10]{School of Science, Huzhou University, Huzhou 313000, Zhejiang, China}
\affil[*]{mengjiao.lyu@nuaa.edu.cn}

\begin{abstract}
    This work introduces a new Control Neural Network (Ctrl.NN) method to
    uncover evidence of exotic quantum state, \textit{i.e.}, the breathing modes
    in 3-$\alpha$ resonant states of $^{12}$C nucleus. We provide the most
    precise microscopic description to date for the $^{12}$C energy spectrum, 
    identify two new exotic breathing states, and uncover strong evidence that
    directly connects the recent experimental observations to the breathing
    modes. The Ctrl.NN method significantly simplifies numerical calculations of
    quantum systems under multiple constraints and offers a new perspective for
    solving the nuclear many-body problem.

\end{abstract}
\begin{document}

\flushbottom
\maketitle
%
%
\thispagestyle{empty}


\section*{Introduction}
In quantum physics, one of the most interesting objectives is to discover the
exotic quantum states, such as the Bose-Einstein condensation that exists in both
cold atom and nuclear physics~\cite{tohsakiAlphaClusterCondensation2001},
formulation of molecular structures of both atomic
nuclei~\cite{PhysRevLett.131.212501} and pentaquark
particles~\cite{PhysRevLett.122.222001}, and the recent discussions on the
breathing-like excitations of nucleons~\cite{PhysRevLett.131.242502} or
$\alpha$-clusters in the nuclear
systems~\cite{li2022multiprobe,li2022investigating}. 

For $^{12}$C nucleus, very rich phenomena arouse in the nuclear excited states,
such as the intriguing formation of stable alpha clusters that constitutes
3-$\alpha$ resonances above the threshold. In particular, the $0_{2}^{+}$ Hoyle
state at resonant energy of $7.65$ MeV is identified as an
$(0S)$-$\alpha$-condensate state, which plays a crucial role in the stellar
nucleosynthesis of Carbon and heavier elements~\cite{hoyle1954nuclear}. The
condensate nature of the Hoyle state could be further revealed by its possible
excitation to the exotic ``breathing-like" state, where one of the
$\alpha$-clusters in $(0S)$ condensation is excited to the higher $(1S)$-orbit
~\cite{PhysRevC.94.044319, PhysRevC.107.044304}. Due to this excitation, the breathing state is characterized by a very large
radius and a giant isoscalar monopole transition strength to the Hoyle state.
The existence of breathing excitation also modifies the temperature dependence
of the 3-$\alpha$ reaction rate at $T_9 \gtrsim 2 $ ($T_9 = T/10^{9}$ K) in
stars and thus potentially revises the stellar evolution and nucleosynthesis
models~\cite{li2022investigating}. 

Recently, a rich energy spectrum has been determined experimentally for the
resonant states of $^{12}$C~\cite{itoh2011candidate, itoh2013nature,
zimmerman2013unambiguous, freer2011evidence, PhysRevC.86.034320,
PhysRevLett.113.012502, li2022multiprobe, li2022investigating}. The $2_{2}^{+}$
and $4_{2}^{+}$ states of the Hoyle band are observed at $10.0$ MeV
\cite{itoh2011candidate, zimmerman2013unambiguous,
li2022multiprobe,li2022investigating} and $13.3$ MeV \cite{freer2011evidence},
which are explained as the rotation of a 3-$\alpha$ cluster structure with an
equilateral triangle shape based on $D_{3h}$ symmetry
\cite{PhysRevLett.113.012502}. However, discussions on the third band of
$^{12}$C is much more limited. Experimentally, a broad $0^{+}$ resonance is
reported at about $10$ MeV~\cite{itoh2011candidate}, which is also reproduced by
recent \textit{ab initio} calculation with a relatively small
radius~\cite{shen2023emergent}. On the other hand, the isoscalar-strength
distribution measured between 8-12 MeV indicates that the broad resonance is
composed by two separate $0_{3}^{+}$ And $0_{4}^{+}$ states that are located at
$9.04$ MeV and $10.56$ MeV, respectively~\cite{itoh2011candidate}.
Theoretically, the $0_{3}^{+}$ state is interpreted as a breathing-like mode of
the Hoyle state~\cite{PhysRevC.94.044319}, where one of the $\alpha$-clusters in
$(0S)$ condensation is excited to the higher $(1S)$-orbit, which is named as
``breathing state". Due to this excitation, the breathing state is characterized
by a very large radius and its giant isoscalar monopole transition strength to
the Hoyle state. In recent experiments, excess monopole strength at $E \approx
9$ MeV has been observed~\cite{li2022multiprobe,li2022investigating}, which
support the existence of giant monopole transition between $0_{3}^{+}$ and
$0_{2}^{+}$ states~\cite{PhysRevC.94.024344,PhysRevC.94.044319,
PhysRevC.107.044304}. For the higher $J$ elements, the $2_{3}^{+}$ state at
$11.1$ MeV has been proposed in early experiment~\cite{AJZENBERGSELOVE19901} and
subsequently reexamined in recent work~\cite{PhysRevC.76.034320}. Additionally,
there is tentative evidence for a broad $2^{+}$ state between $10.5$ and $12$
MeV~\cite{PhysRevC.81.024303}, which is considered as a candidate of the
$2_{3}^{+}$ state. In previous works, this state has not been yet associated to
the breathing states except that it is assigned to the $0_3^{+}$ band in cluster
model calculations~\cite{PhysRevC.92.021302,PhysRevC.99.064327}.

In this work, we propose a new microscopic model of $^{12}$C guided by a
Control Neural Network (Ctrl.NN) that successfully reproduce the existing
experimental spectrum of $^{12}$C above the 3--$\alpha$ threshold energy with
considerable accuracy. We predict the $2_{3}^{+}$ and $4_{3}^{+}$ states as two
new breathing states and provide new evidence that connects current experimental
spectrum to these breathing modes.

\section*{Results}
\label{sec:results and discussion}
\subsection*{Energy spectrum of $^{12}$C }
\begin{figure}[htb]
    \centering
    \includegraphics[width=0.7\linewidth]{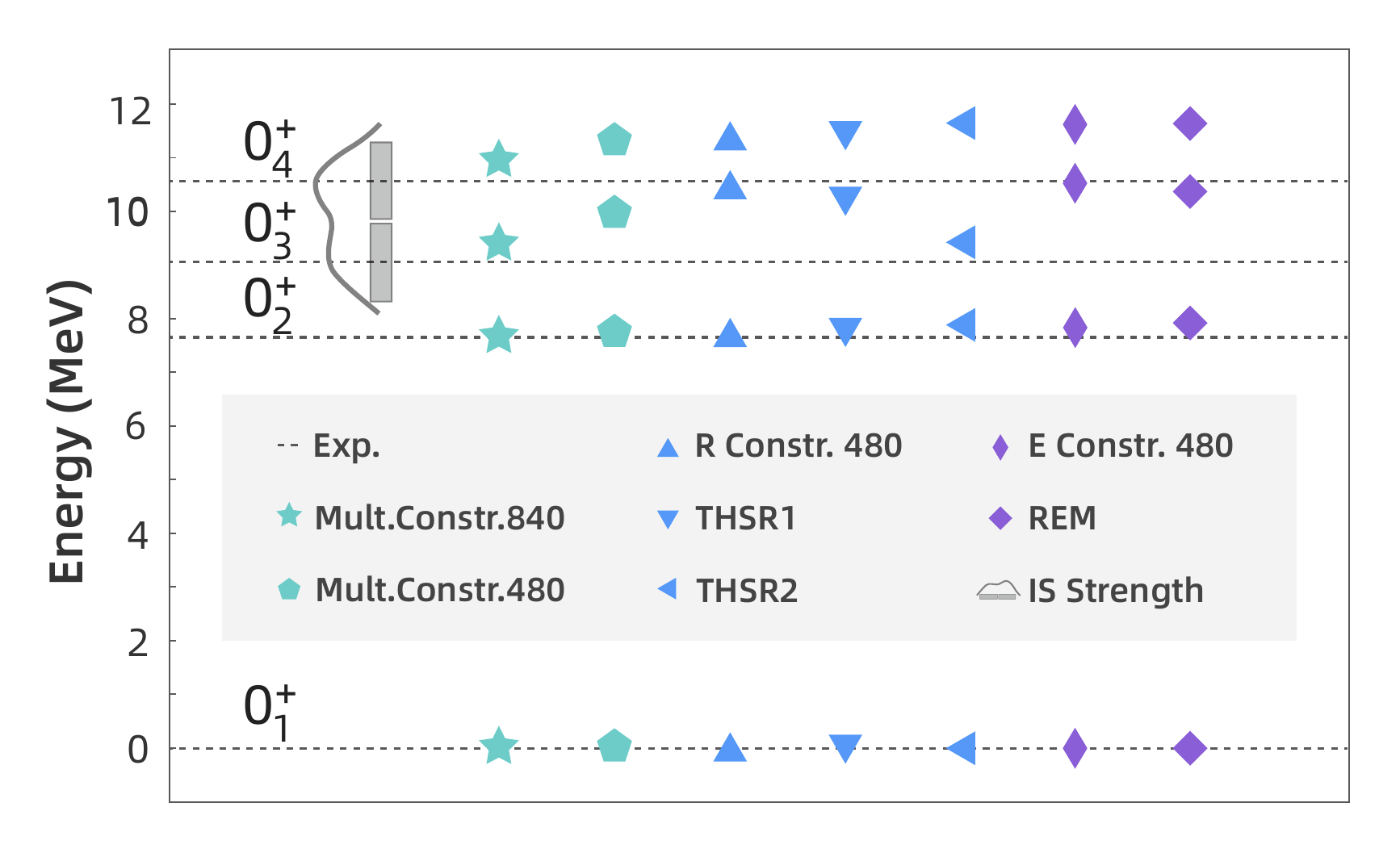}
    \caption{Excitation energies of $0^{+}$ states for $^{12}$C measured from
    the ground state energy. Calculations employing a similar constraint
    strategy are represented with same color. Different microscopic calculation
    results by the THSR1~\cite{PhysRevC.92.021302},
    THSR2~\cite{PhysRevC.94.044319}, REM~\cite{PhysRevC.99.064327} and our
    Ctrl.NN method with multi- or single-objective optimization are compared
    with the experimental
    data~\cite{AJZENBERGSELOVE19901,itoh2011candidate,freer2014hoyle}.    
    The gray curve shows the isoscalar strength distributions from
    Ref~\cite{itoh2011candidate}.}
    \label{fig:Comparison_of_different_constraints}
\end{figure}

We use the Ctrl.NN to evolute $28$ sets of basis wave functions under different
multiple constraint conditions, resulting in a total of $840$ bases, with each
set comprising $30$ cluster wave functions. Then, all these bases are
angular momentum projected and diagonalized to find the energy spectrum and
eigenstates of $^{12}$C. The ground state and Hoyle state energies of the
$^{12}$C nucleus obtained from our calculations are respectively $-89.66$ MeV
and $-81.98$ MeV, consistent with the values calculated in previous
works~\cite{PhysRevC.92.021302,PhysRevC.99.064327}. Our results indicate the
presence of excited $0_{3}^{+}$ and $0_{4}^{+}$ states at $9.37$ MeV and $11.08$
MeV, respectively, which fit very well with the experimental values of $9.04$
MeV and $10.56$ MeV~\cite{itoh2011candidate}. As a benchmark, we compare our
results of the four $0^{+}$ state levels with values from previous
works~\cite{PhysRevC.92.021302,PhysRevC.94.044319,PhysRevC.99.064327} in
Fig.~\ref{fig:Comparison_of_different_constraints}. Additionally, values
calculated with a single constraint in the Ctrl.NN method are shown for
comparison. We have found that the Ctrl.NN method can reproduce results in
previous works such as ``THSR1''~\cite{PhysRevC.92.021302} and
``REM''~\cite{PhysRevC.99.064327}, under single radius or energy constraint,
respectively. By adopting multi-objective optimization in this study, the $0^+$
spectrum shows improved consistency with experimental data compared to the other
works.

\begin{figure}[htb]
    \centering
    \includegraphics[width=0.7\linewidth]{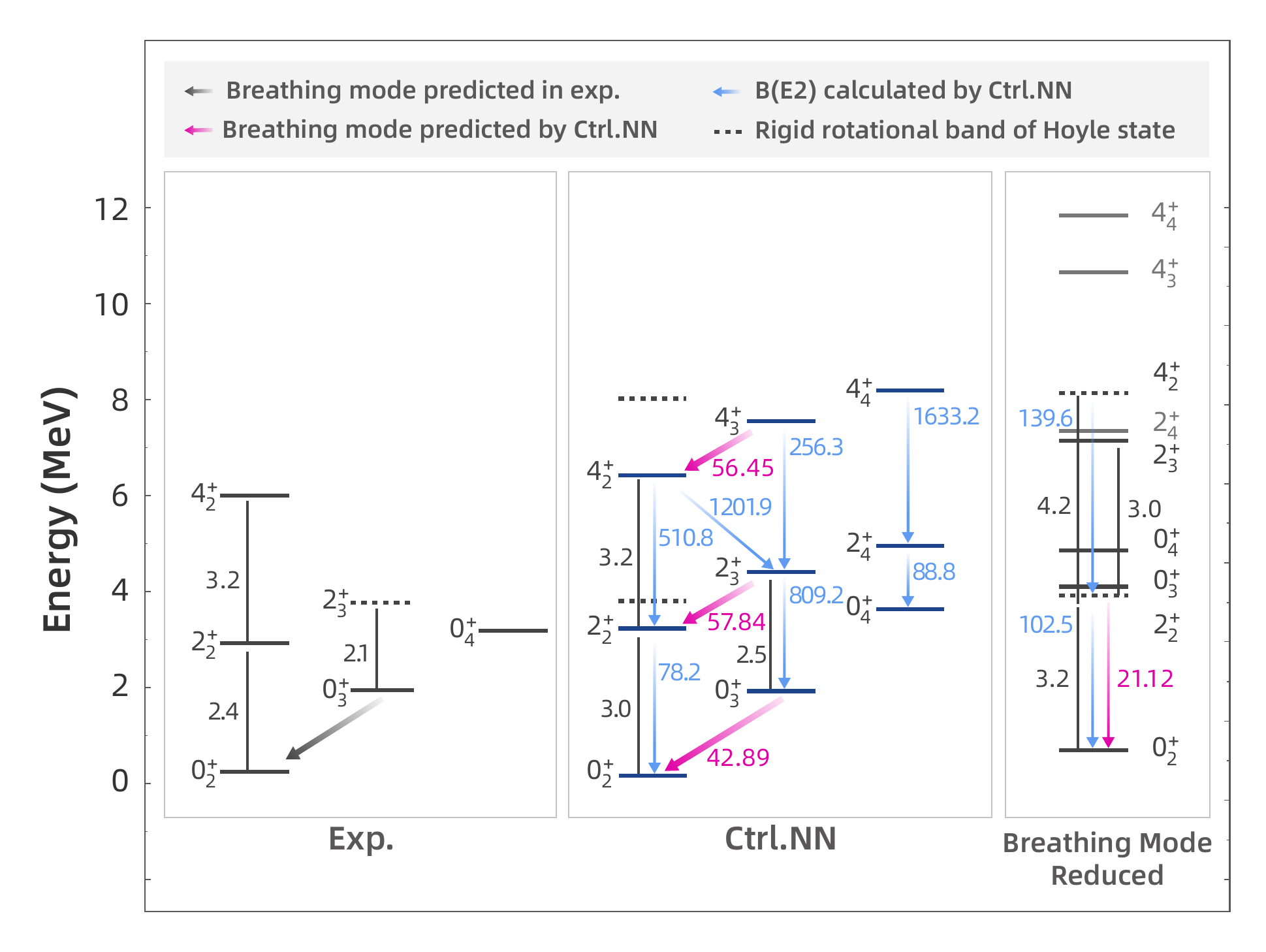}
    \caption{The resonant states energy spectrum of $^{12}$C nucleus. The
    energies were measured from the 3-$\alpha$ threshold energy. The microscopic
    calculation by Ctrl.NN method are compared with experimental results
    \cite{AJZENBERGSELOVE19901,PhysRevC.81.024303,itoh2011candidate,
    itoh2013nature, zimmerman2013unambiguous,freer2011evidence,
    PhysRevC.86.034320,PhysRevLett.113.012502,li2022multiprobe,li2022investigating,
    PhysRevC.76.034320, PhysRevC.81.024303}. Experimental prediction for the
    existence of breathing mode in the Hoyle state comes from
    Refs~\cite{li2022multiprobe,li2022investigating}. Experimental evidence
    supporting the existence of the $2_{3}^{+}$ state is provided in
    Refs.~\cite{PhysRevC.76.034320,PhysRevC.81.024303,AJZENBERGSELOVE19901}.
    }
    \label{fig:Energy-C12}
\end{figure}

We further calculate the complete positive parity energy spectrum of $^{12}$C by
diagonalizing the $840$ bases after angular momentum projection, and compare
with the available experimental observations~\cite{itoh2011candidate,
itoh2013nature, zimmerman2013unambiguous,freer2011evidence,
PhysRevC.86.034320,PhysRevLett.113.012502}, as shown in
Fig.\ref{fig:Energy-C12}. It is noteworthy that the spectrum predicted by the
Ctrl.NN method is in excellent agreement with the existing experimental data for
3-$\alpha$ resonant states, which has not been achieved by microscopic
calculations before. The Hoyle state is successfully reproduced, especially
regarding the energy level spacing between the $0_{3}^{+}$ and $2_{3}^{+}$
states. In previous works, the Hoyle band has been interpreted as a rotational
band characterized by a rigid equilateral triangular intrinsic
structure~\cite{PhysRevLett.113.012502}. In this work, we have calculated the
B(E2) transitions between the energy levels, as shown in
Fig.~\ref{fig:Energy-C12}, and determined the ratio $B(E2;
4_{2}^{+}\rightarrow2_{2}^{+})/B(E2; 2_{2}^{+}\rightarrow0_{2}^{+})$ to be
$6.53$, which exceeds the value of $10/7$. Therefore, our result supports the
non-rigid intrinsic structure in the Hoyle band as already pointed out in
Ref.~\cite{PhysRevC.92.021302, funaki2015cluster}. We also find a large value of
$B(E2; 4_{2}^{+} \rightarrow 2_{3}^{+})$ as $1201.9$ e$^2$fm$^4$, which shows
that the $4_2^+$ state has strong configuration mixing with $4_3^{+}$ state. For
the next $0_3^+$ breathing band, the giant value of $B(E2; 2_{3}^{+} \rightarrow
0_{3}^{+}) = 809.2$ e$^2$fm$^4$ indicates that the $2_3^+$ and $0_3^+$ states
have remarkably similar structures. 
In subsequent analysis, we show that they are breathing counterparts of the
Hoyle band.

\subsection*{New breathing mode of  $^{12}$C} 
To explore the new breathing modes
of $^{12}$C, we calculate the electric monopole transition strengths of the
$0^{+}$, $2^{+}$ and $4^{+}$ states, as shown in Tab~\ref{tab:M_E0}.
\begin{table}[h!]
    \caption{The electric monopole transition strengths $M(E0)$ for $^{12}$C, including the giant values between the breathing states. All units are in efm$^2$.}
    \label{tab:M_E0} 
    \centering
    \begin{tabular}{ccccccccccccccccccc}
        \hline
        \hline
        &Transition &&  $M(E0)$  && Transition &&  $M(E0)$ && Transition &&  $M(E0)$  &   \\
        \hline
        &$0_{2}^{+}\rightarrow 0_{1}^{+}$ &&    6.08 && $2_{2}^{+}\rightarrow 2_{1}^{+}$ &&    5.51 && $4_{2}^{+}\rightarrow 4_{1}^{+}$ &&    2.24&      \\							
        &$0_{3}^{+}\rightarrow 0_{1}^{+}$ &&    3.25  && $2_{3}^{+}\rightarrow 2_{1}^{+}$ &&   4.21 && $4_{3}^{+}\rightarrow 4_{1}^{+}$ &&    3.33&      \\							
        &$0_{4}^{+}\rightarrow 0_{1}^{+}$ &&    3.48 && $2_{4}^{+}\rightarrow 2_{1}^{+}$ &&    0.46 && $4_{4}^{+}\rightarrow 4_{1}^{+}$ &&    2.97&      \\							
       
        &$0_{3}^{+}\rightarrow 0_{2}^{+}$ &&    42.89 && $2_{3}^{+}\rightarrow 2_{2}^{+}$ &&    57.84  && $4_{3}^{+}\rightarrow 4_{2}^{+}$ &&    56.45&      \\							
        &$0_{4}^{+}\rightarrow 0_{2}^{+}$ &&    2.11  && $2_{4}^{+}\rightarrow 2_{2}^{+}$ &&    2.67    && $4_{4}^{+}\rightarrow 4_{2}^{+}$ &&   25.34&      \\							
        &$0_{4}^{+}\rightarrow 0_{3}^{+}$ &&    20.78 && $2_{4}^{+}\rightarrow 2_{3}^{+}$ &&    1.30    && $4_{4}^{+}\rightarrow 4_{3}^{+}$ &&   51.43&      \\							
       
        \hline
        \hline
    \end{tabular}
\end{table}
Here, the most significant transitions are shown as purple arrows in
Fig.\ref{fig:Energy-C12}. Consistent with previous works, we obtain a giant
value for $M(E0;0_{3}^{+}\rightarrow0_{2}^{+})$ as $42.89\ \text{efm}^{2}$,
which shows that the $0_{3}^{+}$ state is indeed the breathing mode of
$0_{2}^{+}$ state. Our prediction for the excitation energy of the $0_3^+$ state is $9.45$ MeV, 
closely matching the experimental value of $E_x \approx 9$ MeV, 
which corresponds to the observation of excess monopole strength~\cite{li2022multiprobe,li2022investigating}. 
Furthermore, we predict
the $2_{3}^{+}$ state as the breathing mode of $2_{2}^{+}$ state, with a giant
transition strength of $M(E0;2_{3}^{+}\rightarrow 2_{2}^{+})=$ $57.84\
\text{efm}^{2}$, which has not yet been discussed in previous works. For the
$M(E0;4_{3}^{+}\rightarrow 4_{2}^{+})$ transition, a large strength of  $56.45\
\text{efm}^{2}$ is obtained, but radius difference between $4_{3}^{+}$  and
$4_{2}^{+}$ states is not significant. Therefore, we consider the $4_{3}^{+}$
state as the ``semi-breathing mode'' of $4_{2}^{+}$. The giant transition
strengths listed above demonstrate the presence of one-to-one correspondence in
the breathing modes between the $0_{3}^{+}$ band and the Hoyle band of
3-$\alpha$ resonances. Thus, we propose to name the $0_{3}^{+}$ band as
``breathing band''.

To provide strong support for the existence of breathing modes in the 3-$\alpha$
resonance, we further scrutinize the substantial contribution to the resonance
spectrum made by the breathing states. This is done by artificially reducing the
breathing mode in the theoretical model space, where basis sets with large
eigenvalues of root-mean-square radii are discarded. Since the breathing states are
characterized by their large radii, the monopole transition strengths
between the Hoyle band and the breathing band are significantly diminished for the
$0^+$, $2^+$, and $4^+$ states as $21.12\ \text{efm}^{2}$, $25.75\ \text{efm}^{2}$,
and $14.66\ \text{efm}^{2}$, respectively. More details are discussed in
Supplementary Note 11. In the reduced model space, the $0_{2}^{+}$ Hoyle state
and the $2_{2}^{+}$ states exhibit relatively minor increment of excitation energy,
as shown in Fig.~\ref{fig:Energy-C12}. However, the excitation energies of
breathing band states, including the $0_{3}^{+}$, $2_{3}^{+}$, and $4_{3}^{+}$,
become significantly higher. Especially, the energy levels of $0_{3}^{+}$ and
$2_{3}^{+}$ states show clear deviation from the experimental values. The
findings indicate that the reduction of the breathing mode results in a
substantial discrepancy between the theoretically energy spectrum and the
experimental observations. In contrast, the remarkable consistency observed in
the full calculation indicates that current experimental observations strongly
support the existence of breathing modes in the 3-$\alpha$ resonance. 

\begin{figure}[htb]
    \centering
    \includegraphics[width=0.6\linewidth]{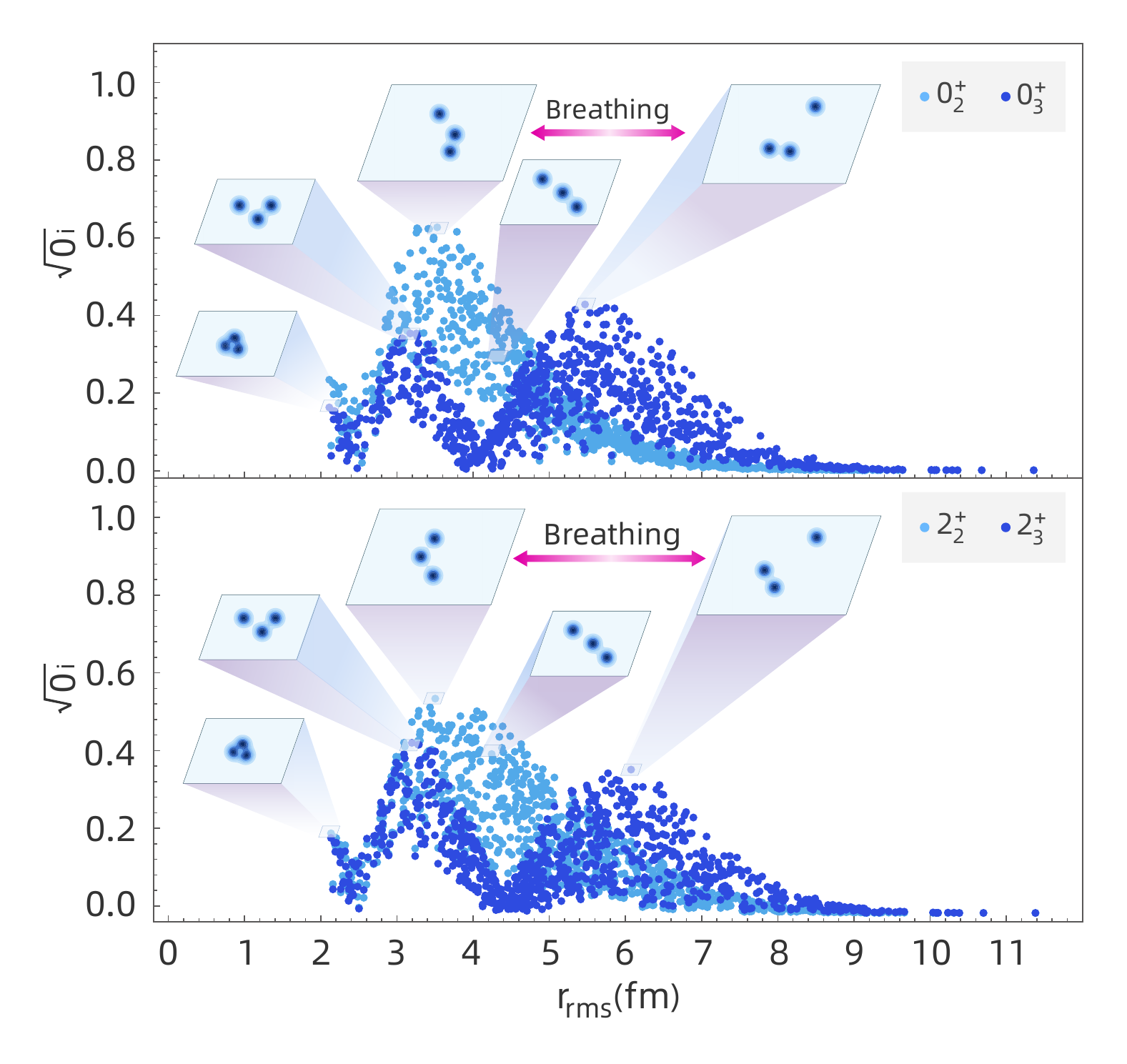}
    \caption{Configuration distribution of $0_{2}^{+}$, $0_{3}^{+}$, $2_{2}^{+}$
    and $2_{3}^{+}$ states. ${O}_{i}$ represents the amplitude. $r_{\text{rms}}$
    represents the root-mean-square radius of each basis. 
    }
    \label{fig:ME0-explain}
\end{figure}

To understand the formation of breathing states, we analyze the contribution
from bases with different configurations to the resonant states. Here, 
$O_{i}$ is defined as the squared overlap (from supplementary material Eq.(32)) between each basis and the total wave function.
We plot the distribution of $\sqrt{O_{i}}$ with respect to the root-mean-square
radius $r_{\text{rms}}$ in Fig.~\ref{fig:ME0-explain}, for the $0^+$ and $2^+$
breathing modes. It is shown that for breathing states $0_{3}^{+}$ and
$0_{2}^{+}$, $\sqrt{O_{i}}$ has a significant distribution over a broad model
space, covering a range of approximately $2{\--}8$ fm, which suggests that
the bases generated by the Ctrl.NN provide a very efficient model space
for the description of giant resonant states of $^{12}$C.
It is found that the darkblue peak at $6$ fm of $^{8}$Be+$^{4}$He configuration plays a
key role in the formation of the breathing state, as it has a large electric
monopole transition strengths to the lightblue peak at $3.5$ fm, which corresponds to
obtuse configurations that contribute most to the $0_{2}^{+}$ and $2_{2}^{+}$
states in the Hoyle band, as shown by the arrows in Fig.~\ref{fig:ME0-explain}.
More detailed information is provided in the Supplementary Note $7{\--}10$.

\section*{Discussion}
In this work, we proposed and performed the microscopic calculations of $^{12}$C,
guided by a Control Neural Network (Ctrl.NN) to uncover the exotic breathing
modes in 3-$\alpha$ resonant states. By imposing multiple constraints on the many-body
wave function via Ctrl.NN, we successfully reproduced the existing
experimental spectrum of $^{12}$C above the 3-$\alpha$ threshold with
considerable accuracy, providing a solid foundation for further analyses.
The rotational band of the Hoyle state was found to be non-rigid by examining
the band level positions and B(E2) transitions. The breathing modes between
the $0_{3}^{+}$ and $0_{2}^{+}$ states were confirmed through significant
monopole transition. Two new breathing modes, \textit{i.e.}, the transitions
between $2_{3}^{+}$ and $2_{2}^{+}$ states and between $4_{3}^{+}$ and
$4_{2}^{+}$ states, were revealed. Artificially reducing the breathing mode
in the theoretical model led to significant deviations in the energy
spectrum prediction for $^{12}$C from the experimental values. In contrast,
the remarkable consistency observed in the full calculation indicated that
current experimental observations strongly support the existence of
breathing modes in the 3-$\alpha$ resonance. This research had been made
feasible by the merits of the new Ctrl.NN method, which significantly
simplifies numerical calculations under multiple constraints and offers a
new perspective for solving the nuclear many-body problem.

\section*{Method}

\subsection*{Control Neural Network Method}
The 3-$\alpha$ resonant states usually exhibit a giant spatial distribution
which lead to significant difficulties for microscopic theories. In recent works, constraint conditions for radius
~\cite{funaki2015cluster, PhysRevC.94.044319, PhysRevC.107.044304, zhou20235},
energy~\cite{PhysRevC.99.064327, zhou2020microscopic, shin2021shape,
zhao2022microscopic}, or
deformation~\cite{suhara2010quadrupole,kanada2012antisymmetrized,
suhara2012cluster, kobayashi2012novel} are usually imposed to limit the model
space, and used independently as formulated in different algorithms. However,
the interplay between the geometric properties, \textit{e.g.} radius and
rotational symmetry, and the dynamical quantities including the energy surface
and derivatives, is essential for the formulation of 3-$\alpha$ resonance.
Hence, for a proper treatment, all these constraints for the resonance wave
functions should be simultaneously involved. Here, we propose a Control Neural
Network (Ctrl.NN) which accepts parameters in the microscopic wave function as
input and learns from multiple constraint conditions with high efficiency in the
evolution of quantum state. Hence we can obtain the trained Ctrl.NN which is capable
to generate proper microscopic wave functions for further calculation. 
This neural network allows general applications for quantum many-body problems by
substituting the wave functions and the constraint targets. 

In this work for the 3-$\alpha$ resonance, we impose multiple constraints
according to the physical properties of the resonant states which are satisfied
by guiding the evolution of cluster wave function via the Ctrl.NN. These include
the spatial extension, rotational symmetry, and energy variation of the cluster
resonant states. In supplementary material, we also show two other applications of control neural
networks for solving quantum many-body problems in both nuclear physics and
condensed matter physics. For energy variational problem, the multiple cooling
method~\cite{myo2023variation}, which can significantly improve the calculations
of the ground and excited states of light atomic nuclei, can be guided by the
Ctrl.NN, such as explained in Sec.\uppercase\expandafter{\romannumeral2} of the
supplementary material for calculations of $^{8-10}$Be nuclei. We also show in
Sec. \uppercase\expandafter{\romannumeral4} of the supplementary material how
the Ctrl.NN can be used to construct cat states targeting at high precision
fidelity and defect density under time boundary conditions for Ising model on
24, 50, and 100 lattices.

\begin{figure}[hbt]
  \centering
  \includegraphics[width=1.0\linewidth]{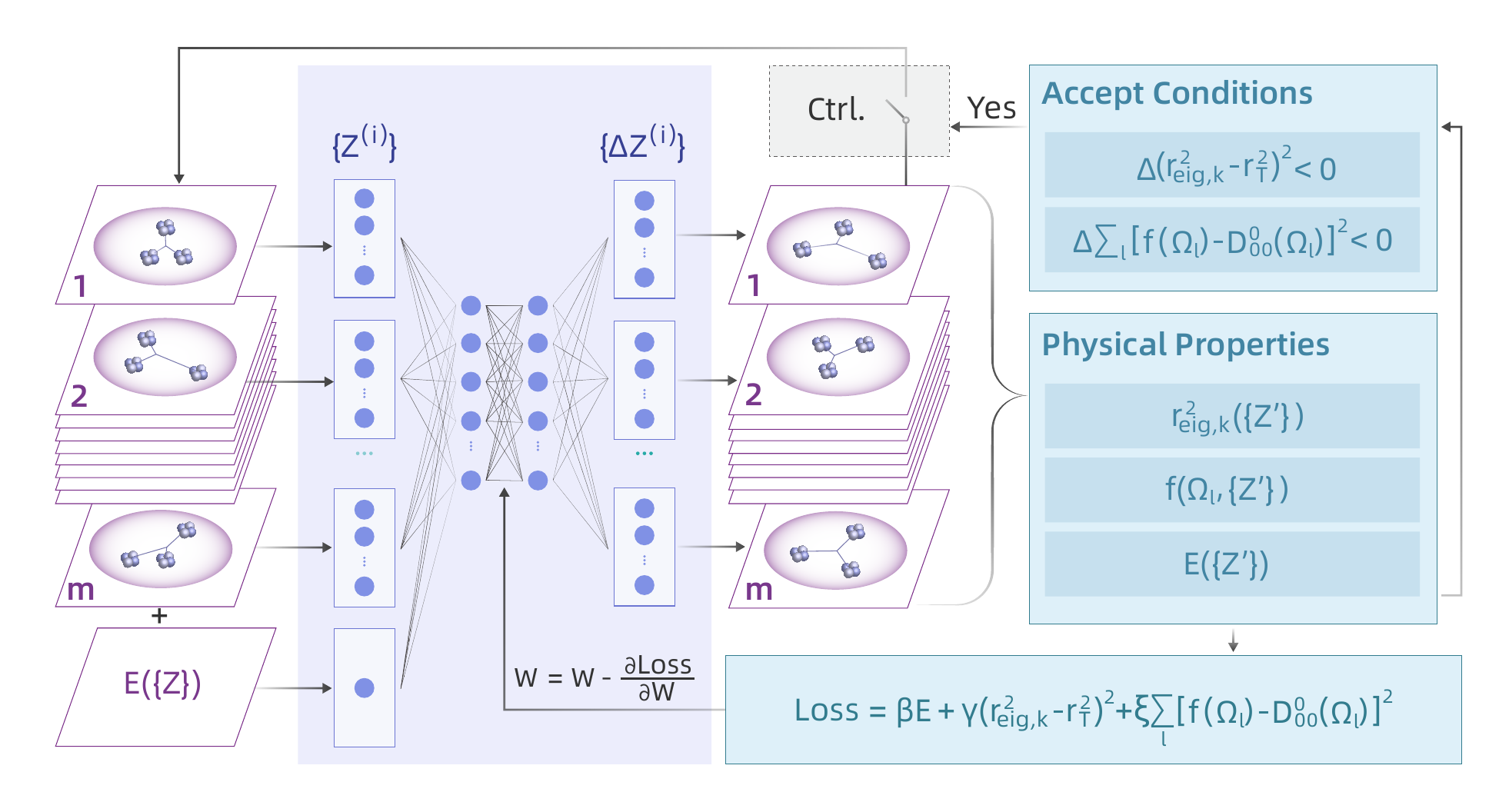}
  \caption{The structures of Control Neural Network. The small slateblue circles
  symbolize the complex nodes of the neural network. The violet modules denote
  the input and output units. The turquoise modules denote the physical quantities
  and algorithms, as introduced in the main text. The update of input is
  controlled by turning on or off the ``Ctrl.'' switch.}
  \label{fig:Structure}
\end{figure}

Here, we explain the fundamental concepts of the Ctrl.NN, with the network
structure shown in Fig.~\ref{fig:Structure}. The more technical explanations with
detailed formulation can be found in Supplementary Note 3. For the 3-$\alpha$
resonant states, we adopt the microscopic cluster wave functions which
explicitly treat antisymmetrization of 12 nucleons. Similar approach are widely
used in recent researches for 5-$\alpha$ systems~\cite{zhou20235}. The detailed
introduction of cluster wave function and effective Hamiltonian can be found in
Supplementary Note 1. The total intrinsic wave function of the 3-$\alpha$ system
is a superposition of the basis wave functions as
\begin{linenomath*}
\begin{equation}
  \begin{aligned}\label{eq:total_wave_function}
      | \Psi \rangle 
      &=
      \sum_{i = 1}^{m}  
      C_{i} \mathcal{A}\{\Phi_{\alpha}(\boldsymbol{Z}^{(i)}_{1}) \Phi_{\alpha}(\boldsymbol{Z}^{(i)}_{2}) \Phi_{\alpha}(\boldsymbol{Z}^{(i)}_{3})\}~.
  \end{aligned}
\end{equation}
\end{linenomath*}
Here, $C_{i}$ is coefficient yielded by the diagonalization, and $\Phi_{\alpha}$
is the microscopic $\alpha$-cluster wave function. The $m$ is the number of
bases in superposition. The $\boldsymbol{Z}^{(i)}_{1}, \boldsymbol{Z}^{(i)}_{2}
, \boldsymbol{Z}^{(i)}_{3}$ are generated coordinates of $\alpha$ clusters, and
we denote them as a set $\{ \boldsymbol{Z}^{(i)} \} = \boldsymbol{Z}^{(i)}_{1},
\boldsymbol{Z}^{(i)}_{2} , \boldsymbol{Z}^{(i)}_{3}$. In order to restore the
rotational symmetry, the total wave function can be obtain after angular
momentum projection technique $\widehat{P}^{J}_{KM}| \Psi \rangle $ (See
Supplementary Note 3 for more details). The input $\boldsymbol{X}$ of the neural
network is expressed as (See Supplementary Note 2 )
\begin{linenomath*}
\begin{equation}
    \begin{aligned}
        \boldsymbol{X}(\{\boldsymbol{Z}\})
        =
        \left[
            \{ \boldsymbol{Z}^{(1)} \} ,\cdots , \{ \boldsymbol{Z}^{(m)} \}, E(\{\boldsymbol{Z}\})
        \right]^{\text{T}},
    \end{aligned}
\end{equation}
\end{linenomath*}
where the intrinsic energy $E(\{\boldsymbol{Z}\})$ of the 3-$\alpha$ system is
obtained diagonalizing the present basis set in the intrinsic frame~\cite{myo2023variation}.

\begin{figure}[htb]
    \centering
    \includegraphics[width=1.0\linewidth]{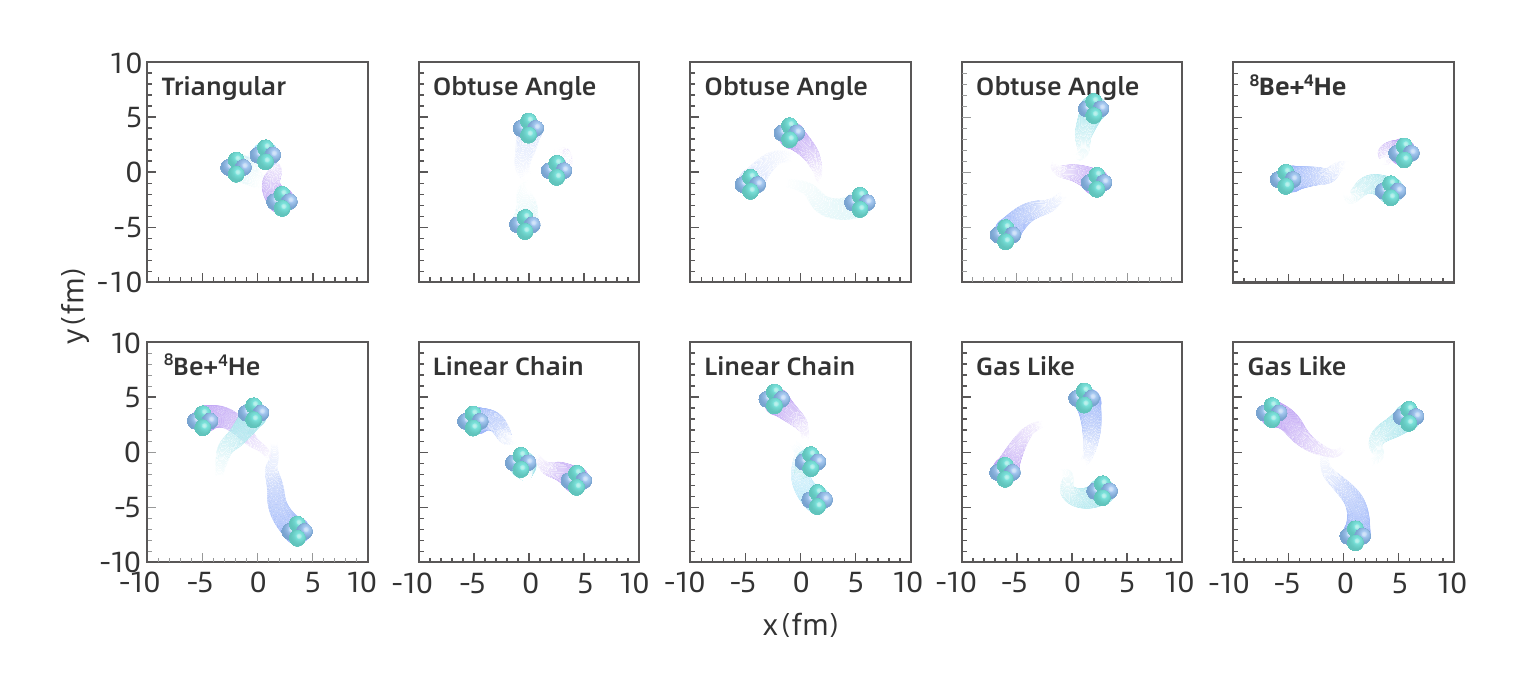}
    \caption{The constrained evolution of basis set guided by the Ctrl.NN. 
    Various three-$\alpha$ configurations emerge from the randomly generated
    initial states. Configurations showen in this figure are selected from a set of 30 bases after evolution. 
    The multiple constraints are introduced in Eq.\ref{eq:total_wave_function} in the main text, where $r^{2}_{T} = 13$ fm$^{2}$, 
    and $J=M=K=0$. (A Movie demonstrating Ctrl.NN guided basis set evolution could be found in Supplementary Movie 1.}
    \label{fig:vector}
  \end{figure}

The output is the amount of change in coordinates $\{\Delta \boldsymbol{Z}^{(i)}
\}$, which is used to form the the new input
$\boldsymbol{X}(\{\boldsymbol{Z}^{\prime}\})$ in the next iteration. All values in
the input, output, and hidden layers are complex numbers. With just-in-time
training, this neural network will gradually optimize the basis set by
predicting the proper coordinate changes to satisfy the three constraints
imposed to the cluster wave function. 
In Fig.\ref{fig:vector}, the constrained evolution of basis set demonstrated, 
where various three-$\alpha$ configurations emerge from the randomly generated
initial states. Detailed discussion for each configuration can be found in Supplementary Note 5.

\subsection*{Multiple constraints}
In each iteration, a new basis set
$\{\boldsymbol{Z}^{\prime}\}$ is obtained and diagonalized to yield the new total
intrinsic wave function $\Psi(\{\boldsymbol{Z}^{\prime}\})$ and physical properties
corresponding to each of the three constraints, including
$E(\{\boldsymbol{Z}^{\prime}\})$,  $r^{2}_{\text{eig},k}(\{\boldsymbol{Z}^{\prime}\})$ and
$f(\Omega_{l})$. Here, $E(\{\boldsymbol{Z}^{\prime}\})$ is the minimum eigenvalue of
the Hamiltonian matrix $ \boldsymbol{H}$ with respect to the new basis set,
$r^{2}_{\text{eig},k}$ is the $k$-th eigenvalue of mean-square radius obtained
by diagonalizing the matrix $\big[\langle \Psi(\{\boldsymbol{Z}^{(i)\prime}\}) |
\widehat{R}^2 | \Psi(\{\boldsymbol{Z}^{(j)\prime}\}) \rangle \big]_{m \times m}$, and
$f(\Omega_{l}, \{\boldsymbol{Z}^{\prime}\}) = \langle \Psi(\{\boldsymbol{Z}^{\prime}\}) |
\widehat{R}(\Omega_{l}) | \Psi(\{\boldsymbol{Z}^{\prime}\}) \rangle$ is a measure of
rotational symmetry through the overlap between the original total intrinsic
wave function $\Psi(\{\boldsymbol{Z}^{\prime}\})$ and the one rotated by the $l$-th
Eular angle $\Omega_{l}$. 
The three properties calculated from the output are constrained for the total
intrinsic wave function by training the neural network with the loss function
formulated as
\begin{linenomath*}
\begin{equation}
    \begin{aligned}\label{eq:Loss}
        \text{Loss}
        = 
        \beta  E 
        + \gamma  (r^2_{\text{eig},k} - r^2_T)^{2}
        + \xi  \sum_{l = 1}^{L} [ f(\Omega_{l}) - D_{MK}^{J}(\Omega_{l})]^{2} , 
    \end{aligned}
\end{equation}
\end{linenomath*}
where $\beta$, $\gamma$, and $\xi$ are the weight parameters of this three
items, with ratio $\beta:\gamma:\xi=1:4:2$. The first term $E$ introduces energy
variation, which reduces the lowest energy with respect to the basis set, and
thus to avoid unphysical states that satisfies only the other two constraint
conditions. The second term $(r^2_{\text{eig},k} - r^2_T)^{2}$ is used to
constrain the spatial extension of nuclear wave function to a target value
$r^{2}_{T}$. This condition is an improved one from the previous works which
selected the eigenstates with eigenvalues smaller than a cut value
$R_{\text{cut}}$ for inclusion in the model
space~\cite{PhysRevC.94.044319,PhysRevC.94.024344}. In this work, we propose
constraining the 2nd to 4th eigenvalues of the mean-square radius to various
target values. This approach allows the corresponding eigenstates, which have
fewer nodal structures, to evolve into states with a relatively larger radius.
This ensures effective coverage of the model space for the resonance states with
substantial spatial extension. The third term $[ f(\Omega_{l}) -
D_{MK}^{J}(\Omega_{l})]^{2}$  constraints the rotational symmetry of the wave
function under rotations, where the Wigner function $D_{MK}^{J}(\Omega_{l})$ is
the target value derived for the eigenstates of angular momentum $J$ with $J_{z}
= M$ or $K$ for bra and ket, respectively~\cite{horiuchi1986cluster}. Using this
constraint, the rotational symmetry is taken into account for the nuclear
resonance states in the intrinsic frame before angular momentum projection, and
thus to limit the model space of the wave function in evolution. The restoration
of this rotational symmetry is discussed in more detail through plotting of the
density distributions (in Supplementary Note 5). 

The merit of the present loss function is the interplay between different
constraints, which collectively limit the model space of many-body wave function
into an essential subset in the evolution, thereby providing the most accurate
description of the physical states after iterations. To speed up the convergence
of wave function evolution, we adopt control conditions for accepting new inputs
from the previous iteration as shown in Fig.\ref{fig:Structure}, where we
discard the basis sets that deteriorate the constraints to the target radius or
rotational symmetry.

\subsection*{Constraint on squared matter radius eigenvalues}

In this section, we explain the constraint for squared matter radius
eigenvalues introduced as the second term $(r^2_{\text{eig},k} - r^2_T)^{2}$ in
$\text{Loss}$ function. We prepare $28$ sets of basis wave functions, each
composed of $30$ cluster wave functions, and impose different control
targets for the 2nd ($k=2$), 3rd ($k=3$), and 4th ($k=4$) eigenvalues of the
$R^{2}$ matrix, where 
\begin{linenomath*}
\begin{equation}\label{eq:RMS}
    \begin{aligned}
        R^{2}_{ij} 
        = \langle \Psi(\{\boldsymbol{Z}^{(i)}\}) |
        \widehat{R}^2 | \Psi(\{\boldsymbol{Z}^{(j)}\}) \rangle~.\\
    \end{aligned}
\end{equation}
\end{linenomath*}
Under this constraint, we evolute each set of bases to a target $r^{2}_{k}$
eigenvalue, and thus prepare a sub model space with proper spatial scale.
Especially, the first few eigenstates of $r^{2}_{k}$ has only small number of
nodes, which ensures effective coverage for the low-lying $0^{+}_{1-4}$ states.
As shown in Supplementary note 5, the largest targets are 8 fm$^{2}$, 16
fm$^{2}$, and  19 fm$^{2}$ for $r^2_{\text{eig},2-4}$, respectively. We note
that this constraint is not used independently, but imposed simultaneously with
the energy and rotational symmetry constraints.


\subsection*{Treatment of resonant states}
In this work, the Hoyle band and the $0_{3}^{+}$ band are all resonant states
located abrove the $\alpha+\alpha+\alpha$ threshold. Resonant states are
time-dependent states which do not break up immediately so that they can be
treated numerically as stationary solutions of the Schr\"odinger equation with
boundary conditions of the outgoing waves. Rigorous solutions to such problems
could be obtained by non-Hermitian quantum mechanics such as complex scaling
methods (CSM)~\cite{moiseyev2011non,myo2014recent}. In the bound state
approximation~\cite{PhysRevC.94.044319,PhysRevC.94.024344,zhou20235,
PhysRevC.92.021302,PhysRevC.99.064327}, the outgoing waves are replaced by
square-integrable waves obtained using variational approach, and validity of the
approximation are verified by correct descriptions of the energy spectra. In the
study of the resonant state of the $^{12}$C nucleus, a large number of bases
with different configurations are often superposed to cover the entire model
space, making it difficult to avoid contamination from the continuum states,
especially when the bases of giant radii are taken into account. However, a
strict constraint for the radius may lead to insufficient model space which can
not reproduce correct energy spectrum. In our study, we demonstrate that this
contradiction is resolved through the imposition of multiple constraints,
whereby the eigenvalues of the squared matter radius are relatively small
($r_{\text{eig}}\le $5 fm), yet the experimental spectrum is accurately
reproduced. Another evidence is the zero contribution to the total wave
functions from the large radius bases. For example, in the
$\sqrt{O_{i}}{\--}r_{\text{rms}}$ distribution as shown in
Fig.~\ref{fig:ME0-explain}, it is fount that $840$ bases cover a broad model
space with radius range of approximately $10$ fm, but the overlap between total
wave function and the bases with radius $r_{\textrm{rms}} > 8$ fm vanishes. As the large
radius bases are essential for the existence continuum states, this distribution
shows that the contamination from continuum is avoided to the largest extent.

\section*{Data availability}
All data relevant to this study are shown in the paper and its Supplementary ﬁle, and more details are available from the corresponding authors.

\section*{Code availability}
Inquiries about the code in this work will be responded to by the corresponding authors.

\bibliography{sample}

\begin{thebibliography}{10}
\urlstyle{rm}
\expandafter\ifx\csname url\endcsname\relax
  \def\url#1{\texttt{#1}}\fi
\expandafter\ifx\csname urlprefix\endcsname\relax\def\urlprefix{URL }\fi
\expandafter\ifx\csname doiprefix\endcsname\relax\def\doiprefix{DOI: }\fi
\providecommand{\bibinfo}[2]{#2}
\providecommand{\eprint}[2][]{\url{#2}}

\bibitem{tohsakiAlphaClusterCondensation2001}
\bibinfo{author}{Tohsaki, A.}, \bibinfo{author}{Horiuchi, H.},
  \bibinfo{author}{Schuck, P.} \& \bibinfo{author}{R{\"o}pke, G.}
\newblock \bibinfo{journal}{\bibinfo{title}{Alpha cluster condensation in
  $^{12}\mathrm{C}$ and $^{16}\mathrm{O}$}}.
\newblock {\emph{\JournalTitle{Phys. Rev. Lett.}}}
  \textbf{\bibinfo{volume}{87}}, \bibinfo{pages}{192501},
  \doiprefix\url{10.1103/PhysRevLett.87.192501} (\bibinfo{year}{2001}).

\bibitem{PhysRevLett.131.212501}
\bibinfo{author}{Li, P.~J.} \emph{et~al.}
\newblock \bibinfo{journal}{\bibinfo{title}{Validation of the
  $^{10}\mathrm{Be}$ ground-state molecular structure using
  $^{10}\mathrm{Be}(p,p\ensuremath{\alpha})^{6}\mathrm{He}$ triple differential
  reaction cross-section measurements}}.
\newblock {\emph{\JournalTitle{Phys. Rev. Lett.}}}
  \textbf{\bibinfo{volume}{131}}, \bibinfo{pages}{212501},
  \doiprefix\url{10.1103/PhysRevLett.131.212501} (\bibinfo{year}{2023}).

\bibitem{PhysRevLett.122.222001}
\bibinfo{author}{Aaij, R.}, \bibinfo{author}{Abell\'an~Beteta, C.},
  \bibinfo{author}{Adeva, B.} \emph{et~al.}
\newblock \bibinfo{journal}{\bibinfo{title}{Observation of a narrow pentaquark
  state, ${P}_{c}(4312{)}^{+}$, and of the two-peak structure of the
  ${P}_{c}(4450{)}^{+}$}}.
\newblock {\emph{\JournalTitle{Phys. Rev. Lett.}}}
  \textbf{\bibinfo{volume}{122}}, \bibinfo{pages}{222001},
  \doiprefix\url{10.1103/PhysRevLett.122.222001} (\bibinfo{year}{2019}).

\bibitem{PhysRevLett.131.242502}
\bibinfo{author}{Michel, N.}, \bibinfo{author}{Nazarewicz, W.} \&
  \bibinfo{author}{P\l{}oszajczak, M.}
\newblock \bibinfo{journal}{\bibinfo{title}{Description of the proton-decaying
  ${0}_{2}^{+}$ resonance of the $\ensuremath{\alpha}$ particle}}.
\newblock {\emph{\JournalTitle{Phys. Rev. Lett.}}}
  \textbf{\bibinfo{volume}{131}}, \bibinfo{pages}{242502},
  \doiprefix\url{10.1103/PhysRevLett.131.242502} (\bibinfo{year}{2023}).

\bibitem{li2022multiprobe}
\bibinfo{author}{Li, K. C.~W.} \emph{et~al.}
\newblock \bibinfo{journal}{\bibinfo{title}{Multiprobe study of excited states
  in $^{12}\mathrm{C}$: Disentangling the sources of monopole strength between
  the energy of the hoyle state and ${E}_{x}=13 \mathrm{MeV}$}}.
\newblock {\emph{\JournalTitle{Phys. Rev. C}}} \textbf{\bibinfo{volume}{105}},
  \bibinfo{pages}{024308}, \doiprefix\url{10.1103/PhysRevC.105.024308}
  (\bibinfo{year}{2022}).

\bibitem{li2022investigating}
\bibinfo{author}{Li, K. C.~W.} \emph{et~al.}
\newblock \bibinfo{journal}{\bibinfo{title}{Investigating the predicted
  breathing-mode excitation of the hoyle state}}.
\newblock {\emph{\JournalTitle{Phys. Lett. B}}} \textbf{\bibinfo{volume}{827}},
  \bibinfo{pages}{136928}, \doiprefix\url{10.1016/j.physletb.2022.136928}
  (\bibinfo{year}{2022}).

\bibitem{hoyle1954nuclear}
\bibinfo{author}{Hoyle, F.}
\newblock \bibinfo{journal}{\bibinfo{title}{On nuclear reactions occuring in
  very hot stars. i. the synthesis of elements from carbon to nickel.}}
\newblock {\emph{\JournalTitle{Astrophys. J. Suppl.}}}
  \textbf{\bibinfo{volume}{1}}, \bibinfo{pages}{121},
  \doiprefix\url{10.1086/190005} (\bibinfo{year}{1954}).

\bibitem{PhysRevC.94.044319}
\bibinfo{author}{Zhou, B.}, \bibinfo{author}{Tohsaki, A.},
  \bibinfo{author}{Horiuchi, H.} \& \bibinfo{author}{Ren, Z.}
\newblock \bibinfo{journal}{\bibinfo{title}{Breathing-like excited state of the
  hoyle state in $^{12}\mathrm{C}$}}.
\newblock {\emph{\JournalTitle{Phys. Rev. C}}} \textbf{\bibinfo{volume}{94}},
  \bibinfo{pages}{044319}, \doiprefix\url{10.1103/PhysRevC.94.044319}
  (\bibinfo{year}{2016}).

\bibitem{PhysRevC.107.044304}
\bibinfo{author}{Takemoto, H.} \emph{et~al.}
\newblock \bibinfo{journal}{\bibinfo{title}{Appearance of the hoyle state and
  its breathing mode in $^{12}\mathrm{C}$ despite strong short-range repulsion
  of the nucleon-nucleon potential}}.
\newblock {\emph{\JournalTitle{Phys. Rev. C}}} \textbf{\bibinfo{volume}{107}},
  \bibinfo{pages}{044304}, \doiprefix\url{10.1103/PhysRevC.107.044304}
  (\bibinfo{year}{2023}).

\bibitem{itoh2011candidate}
\bibinfo{author}{Itoh, M.} \emph{et~al.}
\newblock \bibinfo{journal}{\bibinfo{title}{Candidate for the 2${}^{+}$ excited
  hoyle state at ${E}_{x}\ensuremath{\sim}10 \mathrm{MeV}$ in
  $^{12}\mathrm{C}$}}.
\newblock {\emph{\JournalTitle{Phys. Rev. C}}} \textbf{\bibinfo{volume}{84}},
  \bibinfo{pages}{054308}, \doiprefix\url{10.1103/PhysRevC.84.054308}
  (\bibinfo{year}{2011}).

\bibitem{itoh2013nature}
\bibinfo{author}{Itoh, M.} \emph{et~al.}
\newblock \bibinfo{title}{Nature of 10 $\mathrm{MeV}$ state in
  $^{12}\mathrm{C}$}.
\newblock vol. \bibinfo{volume}{436}, \bibinfo{pages}{012006},
  \doiprefix\url{10.1088/1742-6596/436/1/012006/meta}
  (\bibinfo{organization}{IOP Publishing}, \bibinfo{year}{2013}).

\bibitem{zimmerman2013unambiguous}
\bibinfo{author}{Zimmerman, W.~R.} \emph{et~al.}
\newblock \bibinfo{journal}{\bibinfo{title}{Unambiguous identification of the
  second ${2}^{\mathbf{+}}$ state in $^{12}\mathrm{C}$ and the structure of the
  hoyle state}}.
\newblock {\emph{\JournalTitle{Phys. Rev. Lett.}}}
  \textbf{\bibinfo{volume}{110}}, \bibinfo{pages}{152502},
  \doiprefix\url{10.1103/PhysRevLett.110.152502} (\bibinfo{year}{2013}).

\bibitem{freer2011evidence}
\bibinfo{author}{Freer, M.} \emph{et~al.}
\newblock \bibinfo{journal}{\bibinfo{title}{Evidence for a new
  $^{12}\mathrm{C}$ state at 13.3 $\mathrm{MeV}$}}.
\newblock {\emph{\JournalTitle{Phys. Rev. C}}} \textbf{\bibinfo{volume}{83}},
  \bibinfo{pages}{034314}, \doiprefix\url{10.1103/PhysRevC.83.034314}
  (\bibinfo{year}{2011}).

\bibitem{PhysRevC.86.034320}
\bibinfo{author}{Freer, M.} \emph{et~al.}
\newblock \bibinfo{journal}{\bibinfo{title}{Consistent analysis of the
  2${}^{+}$ excitation of the $^{12}\mathrm{C}$ hoyle state populated in proton
  and \ensuremath{\alpha}-particle inelastic scattering}}.
\newblock {\emph{\JournalTitle{Phys. Rev. C}}} \textbf{\bibinfo{volume}{86}},
  \bibinfo{pages}{034320}, \doiprefix\url{10.1103/PhysRevC.86.034320}
  (\bibinfo{year}{2012}).

\bibitem{PhysRevLett.113.012502}
\bibinfo{author}{Mar\'{\i}n-L\'ambarri, D.~J.} \emph{et~al.}
\newblock \bibinfo{journal}{\bibinfo{title}{Evidence for triangular
  $\mathrm{D}_{3h}$ symmetry in $^{12}\mathrm{C}$}}.
\newblock {\emph{\JournalTitle{Phys. Rev. Lett.}}}
  \textbf{\bibinfo{volume}{113}}, \bibinfo{pages}{012502},
  \doiprefix\url{10.1103/PhysRevLett.113.012502} (\bibinfo{year}{2014}).

\bibitem{shen2023emergent}
\bibinfo{author}{Shen, S.} \emph{et~al.}
\newblock \bibinfo{journal}{\bibinfo{title}{Emergent geometry and duality in
  the carbon nucleus}}.
\newblock {\emph{\JournalTitle{Nat. Commun.}}} \textbf{\bibinfo{volume}{14}},
  \bibinfo{pages}{2777}, \doiprefix\url{10.1038/s41467-023-38391-y}
  (\bibinfo{year}{2023}).

\bibitem{PhysRevC.94.024344}
\bibinfo{author}{Funaki, Y.}
\newblock \bibinfo{journal}{\bibinfo{title}{Monopole excitation of the hoyle
  state and linear-chain state in $^{12}\mathrm{C}$}}.
\newblock {\emph{\JournalTitle{Phys. Rev. C}}} \textbf{\bibinfo{volume}{94}},
  \bibinfo{pages}{024344}, \doiprefix\url{10.1103/PhysRevC.94.024344}
  (\bibinfo{year}{2016}).

\bibitem{AJZENBERGSELOVE19901}
\bibinfo{author}{Ajzenberg-Selove, F.}
\newblock \bibinfo{journal}{\bibinfo{title}{Energy levels of light nuclei
  $\mathrm{A}$ = 11 and 12}}.
\newblock {\emph{\JournalTitle{Nucl. Phys. A}}} \textbf{\bibinfo{volume}{506}},
  \bibinfo{pages}{1--158}, \doiprefix\url{10.1016/0375-9474(90)90271-M}
  (\bibinfo{year}{1990}).

\bibitem{PhysRevC.76.034320}
\bibinfo{author}{Freer, M.} \emph{et~al.}
\newblock \bibinfo{journal}{\bibinfo{title}{Reexamination of the excited states
  of $^{12}\mathrm{C}$}}.
\newblock {\emph{\JournalTitle{Phys. Rev. C}}} \textbf{\bibinfo{volume}{76}},
  \bibinfo{pages}{034320}, \doiprefix\url{10.1103/PhysRevC.76.034320}
  (\bibinfo{year}{2007}).

\bibitem{PhysRevC.81.024303}
\bibinfo{author}{Hyldegaard, S.} \emph{et~al.}
\newblock \bibinfo{journal}{\bibinfo{title}{$r$-matrix analysis of the
  $\ensuremath{\beta}$ decays of $^{12}\mathrm{N}$ and $^{12}\mathrm{B}$}}.
\newblock {\emph{\JournalTitle{Phys. Rev. C}}} \textbf{\bibinfo{volume}{81}},
  \bibinfo{pages}{024303}, \doiprefix\url{10.1103/PhysRevC.81.024303}
  (\bibinfo{year}{2010}).

\bibitem{PhysRevC.92.021302}
\bibinfo{author}{Funaki, Y.}
\newblock \bibinfo{journal}{\bibinfo{title}{Hoyle band and
  $\ensuremath{\alpha}$ condensation in $^{12}\mathrm{C}$}}.
\newblock {\emph{\JournalTitle{Phys. Rev. C}}} \textbf{\bibinfo{volume}{92}},
  \bibinfo{pages}{021302}, \doiprefix\url{10.1103/PhysRevC.92.021302}
  (\bibinfo{year}{2015}).

\bibitem{PhysRevC.99.064327}
\bibinfo{author}{Imai, R.}, \bibinfo{author}{Tada, T.} \&
  \bibinfo{author}{Kimura, M.}
\newblock \bibinfo{journal}{\bibinfo{title}{Real-time evolution method and its
  application to the $3\ensuremath{\alpha}$ cluster system}}.
\newblock {\emph{\JournalTitle{Phys. Rev. C}}} \textbf{\bibinfo{volume}{99}},
  \bibinfo{pages}{064327}, \doiprefix\url{10.1103/PhysRevC.99.064327}
  (\bibinfo{year}{2019}).

\bibitem{freer2014hoyle}
\bibinfo{author}{Freer, M.} \& \bibinfo{author}{Fynbo, H. O.~U.}
\newblock \bibinfo{journal}{\bibinfo{title}{The hoyle state in
  $^{12}\mathrm{C}$}}.
\newblock {\emph{\JournalTitle{Prog. Part. Nucl. Phys.}}}
  \textbf{\bibinfo{volume}{78}}, \bibinfo{pages}{1--23},
  \doiprefix\url{10.1016/j.ppnp.2014.06.001} (\bibinfo{year}{2014}).

\bibitem{funaki2015cluster}
\bibinfo{author}{Funaki, Y.}, \bibinfo{author}{Horiuchi, H.} \&
  \bibinfo{author}{Tohsaki, A.}
\newblock \bibinfo{journal}{\bibinfo{title}{Cluster models from rgm to alpha
  condensation and beyond}}.
\newblock {\emph{\JournalTitle{Prog. Part. Nucl. Phys.}}}
  \textbf{\bibinfo{volume}{82}}, \bibinfo{pages}{78--132},
  \doiprefix\url{10.1016/j.ppnp.2015.01.001} (\bibinfo{year}{2015}).

\bibitem{zhou20235}
\bibinfo{author}{Zhou, B.} \emph{et~al.}
\newblock \bibinfo{journal}{\bibinfo{title}{The 5 $\alpha$ condensate state in
  $^{20}\mathrm{Ne}$}}.
\newblock {\emph{\JournalTitle{Nat. Commun.}}} \textbf{\bibinfo{volume}{14}},
  \bibinfo{pages}{8206}, \doiprefix\url{10.1038/s41467-023-43816-9}
  (\bibinfo{year}{2023}).

\bibitem{zhou2020microscopic}
\bibinfo{author}{Zhou, B.}, \bibinfo{author}{Kimura, M.},
  \bibinfo{author}{Zhao, Q.} \& \bibinfo{author}{Shin, S.-h.}
\newblock \bibinfo{journal}{\bibinfo{title}{Microscopic calculations for
  $\mathrm{Be}$ isotopes within real-time evolution method}}.
\newblock {\emph{\JournalTitle{Eur. Phys. J. A}}}
  \textbf{\bibinfo{volume}{56}}, \bibinfo{pages}{298},
  \doiprefix\url{10.1140/epja/s10050-020-00306-6} (\bibinfo{year}{2020}).

\bibitem{shin2021shape}
\bibinfo{author}{Shin, S.}, \bibinfo{author}{Zhou, B.} \&
  \bibinfo{author}{Kimura, M.}
\newblock \bibinfo{journal}{\bibinfo{title}{Shape of $^{13}\mathrm{C}$ studied
  by the real-time evolution method}}.
\newblock {\emph{\JournalTitle{Phys. Rev. C}}} \textbf{\bibinfo{volume}{103}},
  \bibinfo{pages}{054313}, \doiprefix\url{10.1103/PhysRevC.103.054313}
  (\bibinfo{year}{2021}).

\bibitem{zhao2022microscopic}
\bibinfo{author}{Zhao, Q.}, \bibinfo{author}{Zhou, B.},
  \bibinfo{author}{Kimura, M.}, \bibinfo{author}{Motoki, H.} \&
  \bibinfo{author}{Shin, S.-h.}
\newblock \bibinfo{journal}{\bibinfo{title}{Microscopic calculations of
  $^{6}\mathrm{He}$ and $^{6}\mathrm{Li}$ with real-time evolution method}}.
\newblock {\emph{\JournalTitle{Eur. Phys. J. A}}}
  \textbf{\bibinfo{volume}{58}}, \bibinfo{pages}{25},
  \doiprefix\url{10.1140/epja/s10050-021-00648-9} (\bibinfo{year}{2022}).

\bibitem{suhara2010quadrupole}
\bibinfo{author}{Suhara, T.} \& \bibinfo{author}{Kanada-En'yo, Y.}
\newblock \bibinfo{journal}{\bibinfo{title}{Quadrupole deformation $\beta$ and
  $\gamma$ constraint in a framework of antisymmetrized molecular dynamics}}.
\newblock {\emph{\JournalTitle{Prog. Theor. Phys.}}}
  \textbf{\bibinfo{volume}{123}}, \bibinfo{pages}{303--325},
  \doiprefix\url{10.1143/PTP.123.303} (\bibinfo{year}{2010}).

\bibitem{kanada2012antisymmetrized}
\bibinfo{author}{Kanada-En'yo, Y.}, \bibinfo{author}{Kimura, M.} \&
  \bibinfo{author}{Ono, A.}
\newblock \bibinfo{journal}{\bibinfo{title}{Antisymmetrized molecular dynamics
  and its applications to cluster phenomena}}.
\newblock {\emph{\JournalTitle{Prog. Theor. Exp. Phys.}}}
  \textbf{\bibinfo{volume}{2012}}, \bibinfo{pages}{01A202},
  \doiprefix\url{10.1093/ptep/pts001} (\bibinfo{year}{2012}).

\bibitem{suhara2012cluster}
\bibinfo{author}{Suhara, T.} \& \bibinfo{author}{Kanada-En'yo, Y.}
\newblock \bibinfo{journal}{\bibinfo{title}{Cluster structures in
  $^{11}\mathrm{B}$}}.
\newblock {\emph{\JournalTitle{Phys. Rev. C}}} \textbf{\bibinfo{volume}{85}},
  \bibinfo{pages}{054320}, \doiprefix\url{10.1103/PhysRevC.85.054320}
  (\bibinfo{year}{2012}).

\bibitem{kobayashi2012novel}
\bibinfo{author}{Kobayashi, F.} \& \bibinfo{author}{Kanada-En'yo, Y.}
\newblock \bibinfo{journal}{\bibinfo{title}{Novel cluster states in
  $^{10}\mathrm{Be}$}}.
\newblock {\emph{\JournalTitle{Phys. Rev. C}}} \textbf{\bibinfo{volume}{86}},
  \bibinfo{pages}{064303}, \doiprefix\url{10.1103/PhysRevC.86.064303}
  (\bibinfo{year}{2012}).

\bibitem{myo2023variation}
\bibinfo{author}{Myo, T.} \emph{et~al.}
\newblock \bibinfo{journal}{\bibinfo{title}{Variation of multi-slater
  determinants in antisymmetrized molecular dynamics and its application to
  $^{10}\mathrm{Be}$ with various clustering}}.
\newblock {\emph{\JournalTitle{Phys. Rev. C}}} \textbf{\bibinfo{volume}{108}},
  \bibinfo{pages}{064314}, \doiprefix\url{10.1103/PhysRevC.108.064314}
  (\bibinfo{year}{2023}).

\bibitem{horiuchi1986cluster}
\bibinfo{author}{Horiuchi, H.} \& \bibinfo{author}{Ikeda, K.}
\newblock \bibinfo{journal}{\bibinfo{title}{Cluster model of the nucleus}}.
\newblock {\emph{\JournalTitle{Cluster models and other topics}}}
  \doiprefix\url{10.1142/9789814415453_0001} (\bibinfo{year}{1986}).

\bibitem{moiseyev2011non}
\bibinfo{author}{Moiseyev, N.}
\newblock \emph{\bibinfo{title}{Non-Hermitian quantum mechanics}}
  (\bibinfo{publisher}{Cambridge University Press}, \bibinfo{year}{2011}).

\bibitem{myo2014recent}
\bibinfo{author}{Myo, T.}, \bibinfo{author}{Kikuchi, Y.},
  \bibinfo{author}{Masui, H.} \& \bibinfo{author}{Kat{\=o}, K.}
\newblock \bibinfo{journal}{\bibinfo{title}{Recent development of complex
  scaling method for many-body resonances and continua in light nuclei}}.
\newblock {\emph{\JournalTitle{Progress in Particle and Nuclear Physics}}}
  \textbf{\bibinfo{volume}{79}}, \bibinfo{pages}{1--56},
  \doiprefix\url{10.1016/j.ppnp.2014.08.001} (\bibinfo{year}{2014}).

\end{thebibliography}



\section*{Acknowledgements}
This work is supported by the National Natural Science Foundation of China
(Grants No. 12105141, No. 12035011, No. 11975167), 
by the National Key R\&D Program of China (Contract No. 2023YFA1606503), 
by the Jiangsu Provincial Natural Science Foundation (Grants No. BK20210277), 
by the 2021 Jiangsu Shuangchuang (Mass Innovation and Entrepreneurship) Talent Program (Grants No. JSSCBS20210169), 
by the National Undergraduate Training Program for Innovation and Entrepreneurship (Grants No. 202110287149Y), 
by the JSPS KAKENHI Grant No. JP22K03643, 
and by the JST ERATO Grant No. JPMJER2304, Japan.

\section*{Author contributions}
Z.C. and M.L. proposed the idea of Control Neural Network and formulated the mathematical framework. T.M. proposed the idea of Multiple Cooling for superposed microscopic wave function of atomic nuclei. H.H. proposed the idea of investigation of breathing states. Z.C. performed the coding work and numerical calculation. T.M. and Q.Z. provided important help in formulating the constraint algorithms. M.M. and W.Y. proposed and performed applications of Ctrl.NN into the condensed matter physics in supplementary material. Z.C. and M.L. prepared the manuscript. T.M., H.H., H.T., Z.R., M.I., H.T., N.W., W.Y., and Q.Z. contributed to the discussion of results and were involved in revising the manuscript.

\section*{Competing interests}
The authors declare no competing interests.

\section*{Additional information}






\end{document}